# Byakto Speech: Real-time Long Speech Synthesis with Convolutional Neural Network

## Transfer Learning from English to Bangla

**Zabir Al Nazi**
Brainekt AI Lab
Dhaka, Bangladesh
zabiralnazi@yahoo.com

**Sayed Mohammed Tasmimul Huda**
Dept. of Electronics and Communication Engineering
Khulna University of Engineering and Technology
Khulna, Bangladesh
sm.tasmimulhuda@gmail.com

## Abstract

Speech synthesis is one of the challenging tasks to automate by deep learning, also being a low-resource language there are very few attempts at Bangla speech synthesis. Most of the existing works can't work with anything other than simple Bangla characters script, very short sentences, etc. This work attempts to solve these problems by introducing Byakta, the first-ever open-source deep learning-based bilingual (Bangla and English) text to a speech synthesis system. A speech recogni- tion model based automated scoring metric was also proposed to evaluate the performance of a TTS model. We also introduce a test benchmark dataset for Bangla speech synthesis models for evaluating speech quality. The TTS is available at https://github.com/zabir-nabil/bangla-tts

## 1 Introduction

Bangla speech synthesis is a challenging problem. Most of the existing models are concatenative or parametric in nature. It's hard to represent a rich language like Bangla with such methods. There are many existing approaches that aim to generate Bangla speech.

In [1], authors converted Bangla text to Romanized text based on Bangla graphemes set and by developing a bunch of romanization rules. The proposed system is lightweight and It takes less processing time and produces good under- standable speech.

A phoneme-based Bangla TTs framework was presented in [2]. They kept the dictionary size small which leads to producing a more natural and smooth sound. In their approach, the recorded voice of every alphabet was separated into their element phonemes. This system supports UNICODE input for Bangla Text. In this framework, distortion occurs for large words due to many concatenation points.

[3] described a TTS system for Bangla Language using open-source TTS engine Festival where it used the diphone concatenation approach in its waveform generation phase. The system is able to convert a Unicode encoded Bangla text into human speech and it could be used with any available Screen Reader.

In [4], authors described a method syllabic of unit selection synthesis for converting text into speech for Bangla lan- guage.

A Deep Neural Network-based TTs system with their datasets was presented by researchers from [5]. In this approach, the linguistic features of input text were extracted and then used deep neural networks to map these features with corresponding acoustic features.

[6] proposed multi-speaker acoustic multi-speaker acoustic models using Long Short-Term Memory Recurrent Neural Network (LSTM-RNN) and Hidden Markov Model (HMM) approaches. For that, they used crowdsourcing to collect data from ordinary speakers with small amounts of sentence records.

In [7], authors investigated both phonemes-based and syllable-based approaches of the speech synthesis technique. They showed that syllable-based methods produce higher quality audio than phonemes-based methods.

Most of the above-mentioned approaches work only on a limited Bangla character-set. It's also not easy to generate longer speech and finally the speech quality is very unnatural. In this work, we aim to design a completely deep learning based speech synthesis model that will learn from annotated speech samples to produce arbitrarily long natural- sounding synthetic speech for both Bangla and English character sets.

## 2  Methods

The proposed Bangla speech synthesis system is mainly inspired by [8, 9]. Two deep convolutional networks are used to produce speech. Even though two neural network models are used, as they are fully convolutional it's very fast to train them, and the inference is faster too.

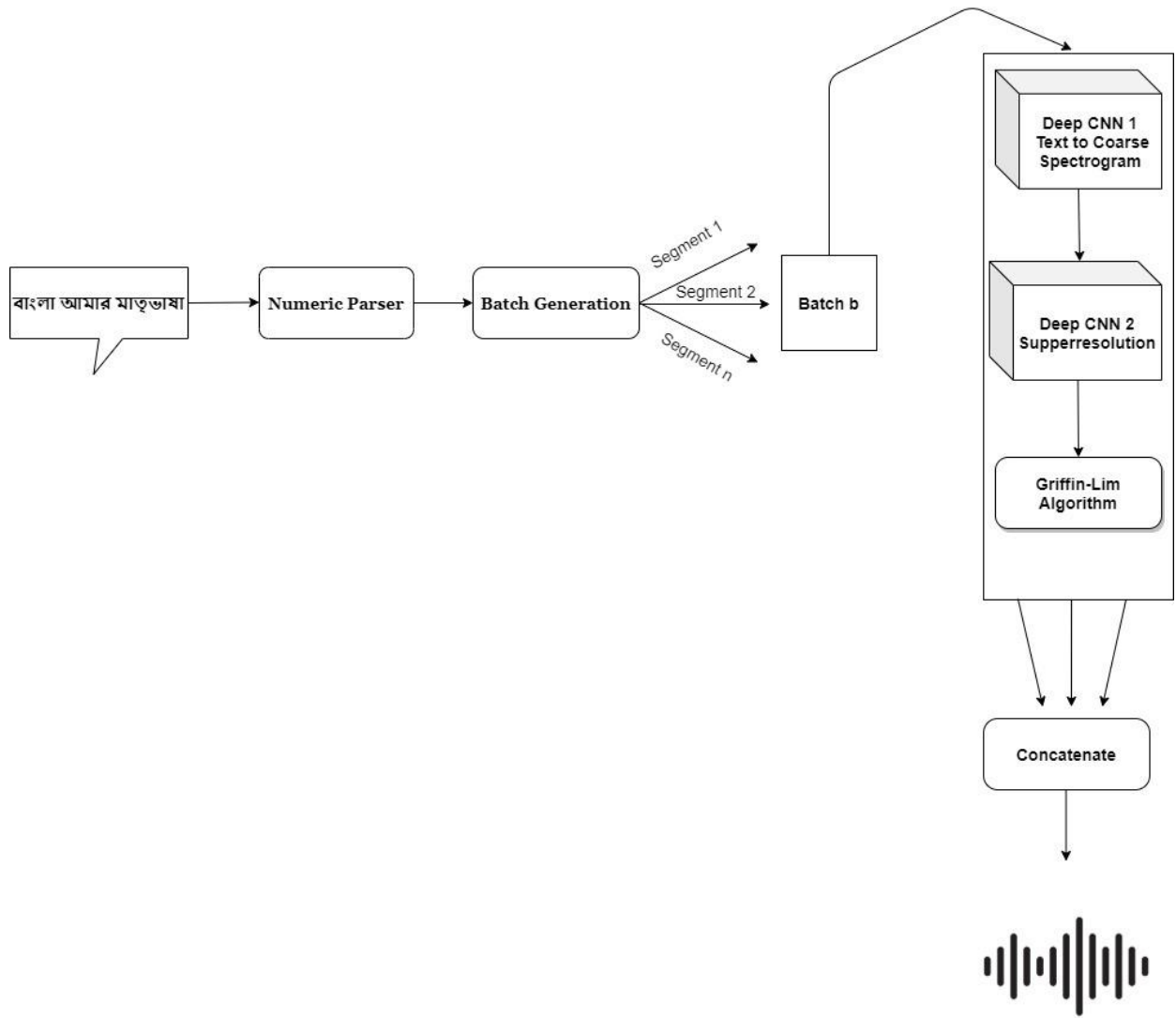

Figure 1: Overall system design of the proposed speech synthesis system

The existing Bangla text-to-speech systems have many limitations. In this work, many modules are designed to solve the problems with existing Text to Speech models available for Bangla.

Table 1: Comparative speech synthesis MOS on Bakta dataset

| Method | MOS |
|---|---|
| Katha Bangla TTS [3, 10] | 2.38 |
| HMM [11] | 3.01 |
| Ours | 3.23 |

## 2.1 Numeric Parsing

One of the major limitations of many Bangla text-to-speech models is not being able to convert numeric values to speech such as phone numbers.                          - most of the TTS will fail in such a case, but it's very important to convert numeric values as they carry much important information. We have designed a numeric parser that can be used not only for our TTS system but for any Bangla TTS system to convert the numeric values to corresponding phonetic representation.

### 2.2 Longer sentence synthesis with batch generation and merge trick

It's challenging to generate longer speech with deep learning models. To overcome this, we have designed a batch generation queue system and merge trick to generate arbitrarily long sentences. We divide any longer sentence into multiple segments, and after the generated speech is retrieved for each segment, they are concatenated to get the final speech.

## 3  Results

In this work, we introduce a test dataset for evaluating Bangla speech synthesis models. The benchmark bakta dataset consists of two lists of sentences, one for only Bangla, and another for Bangla-English. The first list contains 100 sentences from multiple sources such as literature, conversation, drama, song, newspapers. The second list contains 20 sentences₁.

We report comparative mean opinion score on bakta dataset in Table 1. Randomly, 10 sentences were selected from the bakta dataset (80% Bangla, 20% Bangla-English), then 20 subjects submitted quality scores (1 to 5) for three models. Note that, our proposed model can generate speech for both Bangla and Bangla-English sentences whereas others can't.

## 4  Conclusion

Our TTS system is more robust than most of the available Bangla speech synthesis systems. If we integrate more data into the model, it will perform even better. Our Bangla speech synthesis system is the first open-source deep learning-based Bangla TTS system. In the future, we aim to speed up the inference time of the model and improve speech quality by choosing better hyperparameters.

---

1    Bakta dataset is available at https://github.com/zabir-nabil/bangla-tts/tree/master/test_dataset